\documentclass{sbrt2018eng}
\usepackage{comment}
\usepackage{mathrsfs,amsmath}
\usepackage{amsfonts}
\usepackage[utf8]{inputenc}
\usepackage[english, brazil]{babel}

\usepackage{indentfirst,amssymb,newlfont,bm}
\usepackage[usenames,dvipsnames]{color}
\newtheorem{hypothesis}{Hypothesis}
\newtheorem{proposition}{Proposition}
\newtheorem{definition}{Definition}
\usepackage{flushend}
\usepackage{cite}
\usepackage{url}
% ...

\providecommand{\theoremname}{Theorem}

\begin{document}

\title{ND-Wavelets Derived from Anti-symmetric Systems of Isolated Particles using the Determinant of Slater}

\author{H. M. de Oliveira and V. V. Vermehren
\thanks{Statistics Department, Federal University of Pernambuco (UFPE), Recife-PE, Brazil,  E-mail: hmo@de.ufpe.br.
Electrical Engineering Department, University of the State of Amazonas, Manaus, Brazil, victor.vermehren@gmail.com}}

\maketitle

\begin{abstract}
Wavelets are known to be closely related to atomic orbital. A new approach of 2D, 3D and multidimensional wavelet system is proposed from a paralell with anti-symmetric  systems  of  several isolated  particles. The theory of fermionic states is used to generate new \textit{n}-dimensions wavelets, $n\ge2$, by the determinant of Slater. As  pioneering  paper  in  exchanging  formalism between  particle  wave-functions  and  wavelets,  it  opens  some  perspectives for further adaptations derived from the physics of particles in the wavelet analysis scope.
\end{abstract}


\section{Introduction}
\label{sec:introduction}

Wavelet transforms  are important tools in image processing due to their capabilities of multiresolution analysis and image decomposition~\cite{Mallat99,Burrus98}. Wavelet-based image processing is a routinely and largely adopted
approach in texture image decomposition, subband coding, fast image segmentation~\cite{Aujol2006,Antonini92,Vetterli90,Lim90,Kim2003}, even
for 3D~\cite{Taubman94}. There are also plenty of applications of wavelet
in computer graphics~\cite{Muraki93},
\cite{Roerdink99}, \cite{Schroder96}, \cite{Stollnitz96}. 

There is a rooted~\cite{Ash2012}, but not fully explored link between quantum mechanics
and wavelets. The wave nature of light can be deduced from the phenomenon of interference, the photoelectric effect, however, seems to suggest a corpuscular nature of light. Theoretical physicists struggled to include observations like the photoelectric effect and the wave-particle duality into their formulations~\cite{Sully2012}. Erwin Schr\"odinger,
an Austrian physicist, was using advanced mechanics to deal with these phenomena
and developed a equation that relates the space-time in quantum mechanics, in an attempt to get the analog of the classical mechanics. Because wavelets are localized in both time and frequency they offer significant advantages for the
analysis of problems in quantum mechanics.
In this paper, we shed some light on some of these relations.
Rather than seeking at wavelet features on particles or waves,
we adapted some concepts of quantum mechanics to a novel wavelet decomposition for still images.

Initially, we introduce in Section~\ref{sec:motivation} the motivation to build these wavelets, with inspiration in ``Particle Physics'' ~\cite{Eisberg2007,Beiser94}. In Section~\ref{sec:orbital}, we introduce the construction of orbital wavelets for the image decomposition, starting from a single mother wavelet. It is shown this construction actually generates two-dimensional wavelets.  Section~\ref{sec:multidimensional} includes the main lines for an extension to the 3D-case and to higher dimensions.
Finally, the concluding remarks are presented in Section~\ref{sec:concluding}.

\section{Motivation Arising From Particle Physics}
\label{sec:motivation}

Usual wavelet image analysis combines one-dimensional (1D) wavelets  to generate
a two-dimensional (2D) wavelet~\cite{Antonini92,Vetterli90}. This is equivalent to the analysis by the matrix:

\begin{align}
\begin{bmatrix}LL & HL\\
LH & HH
\end{bmatrix},
\label{eq:1}
\end{align}

\hspace{-4mm}where $L$ and $H$ denote low- and high-pass bands, respectively~\cite{Stollnitz95}.

The combinations \emph{LL} and \emph{HH} naturally exhibit a symmetry
(even): $\varphi(x)\cdot\varphi(y)$ = $\varphi(y)\cdot\varphi(x)$,
ditto for $\psi(x)\cdot\psi(y)$ = $\psi(y)\cdot\psi(x)$.
Now, the combination of $\varphi(\cdot)$ and $\psi(\cdot)$
resulting in two different analyzes and non-commutative, namely, $\varphi(x)\cdot\psi(y)$
and $\psi(x)\cdot\varphi(y)$. The combination of $\varphi(\cdot)$
and $\psi(\cdot)$ should result in asymmetry (odd symmetry),
and the exchange of the coordinates $x \leftrightarrow y$ just swap the direction of observation. The combined $\varphi$\_$\psi$
wave should be such that:
\begin{align}
\label{eq:2}
\varphi_{-}\psi(x,y)=-\varphi_{-}\psi(y,x).
\end{align}

The proposal then is to define the orbital combination between $\varphi$
and $\psi$ by the following wave function:
\begin{align}
\varphi_{-}\psi(x,y):=\frac{1}{\sqrt{2}}[\varphi(x)\psi(y)-\psi(x)\varphi(y)], \end{align}
instead of merely $\varphi(x)\cdot\psi(y)$ (\emph{LH}) or $\psi(x)\cdot \varphi(y)$
(\emph{HL}). It should be noted that it now hold Equation (\ref{eq:2}).

Worth nothing that the subtraction of the standard images \emph{LH
- HL} results in exactly the picture of the orbital decomposing, as
well as its negative, on the secondary diagonal of the decomposition
matrix of  Equation \eqref{eq:1}.

\section{Orbital Wavelet for the 2D Case}
\label{sec:orbital}

The wave functions describing electronic orbital can be combined generating
``atomic orbitals''. The ``equivalent''
in the scope of wavelets, also characterized by wave functions, would
be combination of different spatial wavelets~\cite{DeO}. In the
case of particles is typically assumed the combination of anti-symmetric
particles without interaction~\cite{Eisberg2007}. The combination
of $\alpha$ and $\beta$ states should not depend
on which of the particles (1 or 2) is in one of the particular states.
This is called ``exchange degeneracy''. It corresponds
to a probability density of two particles, being one in the alpha
state and another in the beta state not knowing where a particular
state. There are two ways to achieve this~\cite{Duck98}:

\begin{align}
\psi_{S}(r_{1},r_{2}):=\frac{1}{\sqrt{2}}[\psi_{\alpha\beta}(r_{1})+\psi_{\beta\alpha}(r_{2})],
\label{eq:4}
\end{align}
 and\,
\begin{align}
\psi_{A}(r_{1},r_{2}):=\frac{1}{\sqrt{2}}[\psi_{\alpha\beta}(r_{1})-\psi_{\beta\alpha}(r_{2})].
\label{eq:5}
\end{align}
where $r_{1}$ and $r_{2}$ are the positions of particle 1 and 2, and $\alpha$ and $\beta$ are particular states, respectively.\\

\emph{This idea can be used in the decomposition of Equation} (\ref{eq:1}). $\psi_{S}$
\emph{is employed in the main diagonal and} $\psi_{A}$
\emph{in the secondary diagonal.} Interestingly, employing the combination
symmetric diagonally main results in

\begin{align}
LL(x,y):=\frac{1}{\sqrt{2}}[\varphi^{\ast}(x)\varphi(y)+\varphi^{\ast}(y)\varphi(x)],
\end{align}
and\,
\begin{align}
HH(x,y):=\frac{1}{\sqrt{2}}[\psi^{\ast}(x)\psi(y)+\psi^{\ast}(y)\psi(x)].
\end{align}

This definition allows, in particular, analyzing images using continuous
complex wavelets (see also~\cite{Kingsbury99}). In the case of real
wavelets, simplification collapses to the usual $\varphi(x) \cdot \varphi(y)$
or $\psi(x)\cdot\psi(y)$. For the sake of simplicity, we often drop the variables $x$ and $y$ and denote by (\emph{LL}) and (\emph{HH}) as introduced in Equation \eqref{eq:1}. The orbital-based 2D analysis is shown in Appendix.

The initial proposal for ``combination''
of two wavelets was adopted for image analysis (2D) considering the
same orthogonal mother wavelet, but at different scales of multiresolution.
The two wavelets are $\psi_{a_{1},b}$(.) and $\psi_{a_{2},b}$(.)
and the approach for simultaneous analysis in two different scales
corresponds to the following set-up: Consider a 1{D} orthogonal~\cite{Lawton91,
Maab96, DeO} wavelet mother $\psi(x)$ equipped
with her daughters $\{\psi_{a,b}(x)\}_{a\neq0,b\in R}$.
\vspace{0.5 cm}
\begin{definition} 
\label{def}
The function \mbox{2D}-orbital at the scales $\{a_{1},a_{2}\}$
is defined by:

\begin{align}
\psi_{A}(x,y):=\frac{1}{\sqrt{2}}\operatorname{det}\begin{bmatrix}\psi_{a_{1},b}^{*}(x) & \psi_{a_{1},b}^{*}(y)\\
\psi_{a_{2},b}(x) & \psi_{a_{2},b}(y)
\end{bmatrix},
\end{align}
\end{definition}

\hspace{-4mm}what can be rewritten as

\begin{align}\begin{split}
\psi_{A}(x,y)=\frac{1}{\sqrt{2|a_{1}||a_{2}|}}\psi^{\ast}\left(\frac{x-b}{a_{1}}\right)\psi\left(\frac{y-b}{a_{2}}\right)\\
&\hspace{-49mm}-\frac{1}{\sqrt{2|a_{1}||a_{2}|}}\psi\left(\frac{x-b}{a_{2}}\right)\psi^{\ast}\left(\frac{y-b}{a_{1}}\right).
\end{split}\end{align}

The condition $a_{1}\neq a_{2}$ eliminates the degeneration into
$\psi_{A}(x,y)=0$. This is to some extent in connection to the
Pauli Exclusion Principle~\cite{Duck98}. It state that with a single-valued many-particle wavefunction is equivalent to requiring the wavefunction to be antisymmetric. An antisymmetric two-particle state is represented as a sum of states in which one particle is in state $\alpha$ and the other in state $\beta$. Besides, the relationship $\psi_{A}(y,x)=-\psi_{A}(x,y)$ ensures the desired asymmetry. Here, we use the same wavelet-mother, but on
different scales. It will be seen that the orthogonality requirements
correspond to the ``dual'' of particles
without interaction. If the wavelets are orthogonal on any two scales,
one can perform a decomposition of an image ``simultaneously''
in both scales. The decomposition 2D sated in  Definition 1 results in
a strict 2D-wavelet.
\vspace{0.5 cm}
\begin{hypothesis} If the wavelets $\{\psi_{a,b}(t)\}$ are orthogonal,
then the inner product $\langle\psi_{a_{1},b},\psi_{a_{2},b}\rangle$=
0 and the following integrals cancel out $\forall a_{1}\neq a_{2}$:

\begin{align}
\intop_{-\infty}^{\infty}\psi_{a_{1},b}(x)\cdot\psi_{a_{2},b}^{*}(x)\mathrm{d}x=\intop_{-\infty}^{\infty}\psi_{a_{1},b}^{*}(x)\cdot\psi_{a_{2},b}(x)\mathrm{d}x=0.
\end{align}

It is also noteworthy that

\begin{align}
\langle\psi_{a_{1},b},\psi_{a_{2},b}\rangle^{*}=\langle\psi_{a_{2},b},\psi_{a_{1},b}\rangle.
\end{align}
\end{hypothesis}
\vspace{0.5 cm}
\begin{proposition}
The previously defined 2D-orbital function has oscillatory behavior
satisfying the following properties:
\vspace{0.3 cm}
\begin{enumerate}
\item
$\intop_{-\infty}^{\infty}\psi_{A}(x,y)\mathrm{d}x=0$,
\item
$\intop_{-\infty}^{\infty}\psi_{A}(x,y)\mathrm{d}y=0$,
\item
$\intop_{-\infty}^{\infty}\intop_{-\infty}^{\infty}\psi_{A}(x,y)\mathrm{d}x\mathrm{d}y=0$.
\end{enumerate}
\end{proposition}
\vspace{0.5 cm}
\begin{proof}
It follows that

\begin{align}
\intop_{-\infty}^{\infty}\psi_{A}(x,y)\mathrm{d}x=\frac{1}{\sqrt{2}}\psi_{a_{2},b}(y)\cdot\overline{\psi_{a_{1},b}^{\ast}(x)}
-\frac{1}{\sqrt{2}}\overline{\psi_{a_{2},b}(x)}\cdot\psi_{a_{1},b}^{\ast}(y),
\end{align}

\hspace{-4mm}where
\begin{align}
\overline{\psi_{a,b}(x)}=\intop_{-\infty}^{\infty}\psi_{a,b}(x)\mathrm{d}x.
\end{align}

\hspace{-4mm}Therefore item 1 derives from the fact that $\psi_{a,b}(x)$,
$a=\{a_{1},a_{2}\}$ be individual wavelets. Demonstration of item {2}
is similar, considering that

\begin{align}
\intop_{-\infty}^{\infty}\psi_{A}(x,y)\mathrm{d}y=\\
\frac{1}{\sqrt{2}}\overline{\psi_{a_{2},b}(y)}\cdot\psi_{a_{1},b}^{\ast}(x)-\frac{1}{\sqrt{2}}\psi_{a_{2},b}(x)\cdot\overline{\psi_{a_{1},b}^{\ast}(y)}.
\end{align}

Now the condition $\intop_{-\infty}^{\infty}\intop_{-\infty}^{\infty}\psi_{A}(x,y)\mathrm{d}x\mathrm{d}y=0$
follows from Fubini's theorem~\cite{DeFig2000}, regardless of the
order of integration. \end{proof}
\vspace{0.5 cm}
\begin{proposition}
The 2D-orbital functions have normalized energy. \end{proposition}
\indent 
\begin{proof}
Computing $|\psi_{A}(x,y)|^{^{2}}=\psi_{A}(x,y)\cdot\psi_{A}^{\ast}(x,y)$,
we arrive at:

\begin{align}
|\psi_{A}(x,y)|^{^{2}}=\frac{1}{2}|\psi_{a_{1},b}(x)|^{^{2}}\cdot|\psi_{a_{2},b}(y)|^{^{2}} \times|\psi_{a_{1},b}(y)|^{^{2}}\cdot|\psi_{a_{2},b}(x)|^{^{2}}
\end{align}

\hspace{-4mm}and therefore

\begin{align}
\intop_{-\infty}^{\infty}\intop_{-\infty}^{\infty}|\psi_{A}(x,y)|^{^{2}}\mathrm{d}x\mathrm{d}y=1.
\end{align}

\end{proof}

If, by hypothesis $\psi_{a_{1},b}(t)\perp\psi_{a_{2},b}(t)$, then
all cross terms are void, concluding the proof. It is possible (more
easily) to combine orthogonal 1D-wavelets and use them to build a
new 2D-wavelet. 
\vspace{0.3 cm}
\begin{proposition}
The 2D-orbital function is a 2D wavelet.\end{proposition}
\vspace{0.3 cm}
\indent 
\begin{proof}
Starting from $\psi(t)\leftrightarrow\Psi(w) $ and the fact that the admissibility condition holds~\cite{Moretin99,Burrus98},\\
\noindent
$C_{\psi}:=\intop_{-\infty}^{\infty}\frac{|\Psi(w)|^{^{2}}}{|w|}dw<+\infty$, and similarly for their daughter wavelets $\psi_{a,b}(t)\leftrightarrow\Psi_{a,b}(w)$
results in $\intop_{-\infty}^{\infty}\frac{|\Psi_{a,b}(w)|^{^{2}}}{|w|}dw<\infty$,
since $\Psi_{a,b}(w)=\sqrt{|a|}\Psi(aw)e^{^{-jwb}}$ \\ \cite{DeO2}.
Let us now evaluate the condition for the 2D case. If the Fourier
transform pair $\psi_{A}(x,y)\leftrightarrow\Psi_{A}(u,v)$ do exist,
the 2D-spectrum of $\psi_{A}~$ can be computed in terms of the 1D-spectrum
of $\psi$:

\begin{align}
\Psi_{A}(u,v)=\frac{\sqrt{|a_{1}a_{2}|}}{\sqrt{2}}\left[\Psi(a_{1}u)\Psi^{\ast}(a_{2}v)-\Psi(a_{2}u)\Psi^{\ast}(a_{1}v)\right].
\end{align}

From the generalized Parseval-Plancherel energy theorem~\cite{DeFig2000,
Burrus98}, the cross-terms vanish due to the orthogonality, and
\begin{equation}
\begin{split}
|\Psi_{A}(u,v)|^{^{2}}=\\
\frac{|a_{1}a_{2}|}{2}|\Psi(a_{1}u)|^{^{2}}|\Psi(a_{2}v)|^{^{2}} +\frac{|a_{1}a_{2}|}{2}|\Psi(a_{2}u)|^{^{2}}|\Psi(a_{1}v)|^{^{2}}.
\end{split}
\end{equation}
Letting
\begin{align}
C_{\psi_{A}}:=\intop\intop_{-\infty}^{\infty}\frac{|\Psi_{A}(u,v)|^{^{2}}}{|u|\cdot|v|}\mathrm{d}u\mathrm{d}v,
\end{align}
the proof that $C_{\psi_{A}}<+\infty$ follows directly from the marginal admission conditions of the 1D daughter-wavelets. 
\end{proof}

\section{The 3D-Orbital Wavelets}
\label{sec:multidimensional}

For extension of the results in the 3D case (or even higher dimensions),
one might consider the following definition inspired by the ``Slater
determinant''~\cite{Slater29}, which is used for antisymmetric
systems of several isolated particles (fermionic state).

\begin{definition}
The 3D-orbital function at three distinct scales
$\{a_{1},a_{2},a_{3}\}$ is given by
\begin{align}
\psi_{A}(x,y,z)
=
\frac{1}{\sqrt{3!}} \cdot
\operatorname{det}
\begin{bmatrix}
\psi_{a_{1},b}^{^{\ast}}(x) & \psi_{a_{1},b}^{^{\ast}}(y) & \psi_{a_{1},b}^{^{\ast}}(z)\\
\psi_{a_{2},b}(x) & \psi_{a_{2},b}(y) & \psi_{a_{2},b}(z)\\
\psi_{a_{3},b}^{^{\ast}}(x) & \psi_{a_{3},b}^{^{\ast}}(y) & \psi_{a_{3},b}^{^{\ast}}(z)
.
\end{bmatrix}
\end{align}
\end{definition}

The general case follows the same lines. Again, ensuring the orthogonality
of the 1D wavelet chosen as the starting point is an essential statement. Further generalizations can also be done.

\section{Linear Combination of wavelets as LCAO}
\label{sec:LCAO}

This paper presented a way to combine orthogonal wavelets, Equation \eqref{eq:4} and \eqref{eq:5}, using a method similar to the LCAO (Linear Combination of Atomic Orbitals) approach ~\cite{Manh87}. When there is no orthogonality, one possible solution is to consider the quantity $S\neq0$, called orbital overlap (or recoating), which defined by:

\begin{align}
S:=\intop_{-\infty}^{\infty}\psi_{a_{1},b}(x)\cdot\psi_{a_{2},b}^{*}(x)\mathrm{d}x
\end{align}

In this cases Equation \eqref{eq:4} and Equation \eqref{eq:5} are replaced by

\begin{align}
\widetilde{\psi_{S}}(r_{1},r_{2}):=\frac{1}{\sqrt{2(1+S)}}[\psi_{\alpha\beta}(r_{1})+\psi_{\beta\alpha}(r_{2})],
\end{align}
 and\,
\begin{align}
\widetilde{\psi_{A}}(r_{1},r_{2}):=\frac{1}{\sqrt{2(1-S)}}[\psi_{\alpha\beta}(r_{1})-\psi_{\beta\alpha}(r_{2})].
\end{align}
This can clearly be put as a generalization on Definition 1, but now in the wavelet framework. Since LCAO is larged and successful used in molecule studies, this similar wavelet approach can has a potential use.  

\section{Concluding Remarks}
\label{sec:concluding}

Even prospective and introductory, the main ideas presented here can be explored, taking advantage of the cornucopia of tools used in particle physics and atomic orbitals theory. This paper offers an original and general approach for image
decomposition engendered by asymmetric orthogonal wavelets, which
allows much room, somewhat akin to the extension from wavelet to wavelet
packets. Despite the focus being essentially on still image, this
approach allows a fully scalable multimedia decomposition. It remains
to be investigated the potential of this approach in image compressing
\cite{DeVore95}, in 3D processing and scalable coding for multimedia
schemes~\cite{Ohm2004}. Applications in other scenarios such as wavelet-based watermarking~\cite{JHU2002} or steganography~\cite{Carrion2008} also deserve an investigation. As pioneering paper in exchanging formalism between particle wave-functions and wavelets, it opens new perspectives for adaptations derived from the quantum chemistry in the
wavelet analysis scope.

\section{Acknowledgement}
The authors are grateful to Dr. Renato Cintra (UFPE Department of Statistics) who has actively contributed to the development of the main concepts related to the 2D-wavelet model linked to atomic orbitals. They also thanks to Dr. R. Ospina (UFPE Department of Statistics) for valuable support on \LaTeX ~issues.

\appendix
\section{}
\noindent
A general alternative formulation of standard image (2D) wavelet decomposition can be carried out by the following

\begin{equation*}
    \begin{bmatrix}LL & HL\\
LH & HH
\end{bmatrix},
\end{equation*}

\hspace{-4mm}where $L$ and $H$ denote low- and high-pass bands, respectively.

\begin{equation*}
    LL=\varphi\_\varphi_{S}(x,y):=\frac{1}{\sqrt{2}}\operatorname{det}\begin{bmatrix}\varphi^{*}(x) & \varphi^{*}(y)\\
-\varphi(x) & \varphi(y)
\end{bmatrix}.
\end{equation*}

\begin{equation*}
    HH=\psi\_\psi_{S}(x,y):=\frac{1}{\sqrt{2}}\operatorname{det}\begin{bmatrix}\psi^{*}(x) & \psi^{*}(y)\\
-\psi(x) & \psi(y)
\end{bmatrix}.
\end{equation*}

\begin{equation*}
    LH=\varphi\_\psi_{A}(x,y):=\frac{1}{\sqrt{2}}\operatorname{det}\begin{bmatrix}\varphi^{*}(x) & \varphi^{*}(y)\\
\psi(x) & \psi(y)
\end{bmatrix}.
\end{equation*}

\begin{equation*}
    HL=\psi\_\varphi_{A}(x,y):=\frac{1}{\sqrt{2}}\operatorname{det}\begin{bmatrix}\psi^{*}(x) & \psi^{*}(y)\\
\varphi(x) & \varphi(y)
\end{bmatrix}.
\end{equation*}
The symmetries involved are:\\
$\varphi\_\varphi_{S}(x,y)=\varphi\_\varphi_{S}(y,x)$ and 
$\psi\_\psi_{S}(x,y)=\psi\_\psi_{S}(y,x)$,\\ whereas the antisymmetries are $\varphi\_\psi_{A}(x,y)=-\varphi\_\psi_{A}(y,x)$ and $\psi\_\varphi_{A}(x,y)=-\psi\_\varphi_{A}(y,x)$.
\end{document}